\documentclass[conference]{IEEEtran}

\usepackage{times}
\usepackage{latex8}
\usepackage{graphicx}
\usepackage{float}
\usepackage{chngpage}
\usepackage{xspace}
\usepackage{array}
\usepackage[flushleft]{threeparttable}
\PassOptionsToPackage{hyphens}{url}\usepackage{hyperref}
\usepackage{bookmark}
\usepackage{listings}
\usepackage[labelfont=bf,font=bf]{caption}

\newfloat{listing}{thp}{lop}
 \floatname{listing}{Listing}

\newcommand{\cwb}{\textsc{CWB}\xspace}

\newcommand{\ie}{\emph{i.e.},\xspace}
\newcommand{\eg}{\emph{e.g.},\xspace}

\begin{document}

\title{Cloud WorkBench -- Infrastructure-as-Code Based Cloud Benchmarking}

\author{Joel Scheuner, Philipp Leitner, J\"urgen Cito, Harald Gall\\
s.e.a.l. -- software evolution \& architecture lab \\
University of Zurich,
Switzerland \\
             \{lastname\}@ifi.uzh.ch
}

\maketitle

\begin{abstract}
To optimally deploy their applications, users of Infrastructure-as-a-Service clouds are required to evaluate the costs and performance of different combinations of cloud configurations to find out which combination provides the best service level for their specific application. Unfortunately, benchmarking cloud services is cumbersome and error-prone. In this paper, we propose an architecture and concrete implementation of a cloud benchmarking Web service, which fosters the definition of reusable and representative benchmarks. In distinction to existing work, our system is based on the notion of Infrastructure-as-Code, which is a state of the art concept to define IT infrastructure in a reproducible, well-defined, and testable way. We demonstrate our system based on an illustrative case study, in which we measure and compare the disk IO speeds of different instance and storage types in Amazon EC2.
\end{abstract}

\section{Introduction}

The idea of cloud computing \cite{Buyya:2011le,buyya:09} is a new paradigm, that has the potential to fundamentally change the IT industry. In cloud computing, resources, such as virtual machines (VMs), programming environments, or entire application services, are acquired on a pay-per-use basis. In the IaaS model of cloud computing, \emph{"processing, storage, networks, and other fundamental computing resources"} \cite{Mell:2011tg} are acquired on a pay-per-use basis, most commonly in the form of virtual machines (VMs). The functional similarities of these services are contrasted by significant variations in non-functional properties. Service performance not only varies between providers, as studies listed in \cite{Farley:2012vh} show, but also for services exhibiting the same specification \cite{Gillam:2013if}. Under these conditions, software engineers obtain the best results for service selection in terms of accuracy and relevance by running the raw (\ie real world) application in the cloud, as shown for instance in \cite{borhani:14,krishnappa:12}. However, systematic benchmarking (\ie performance testing) of actual applications is practically hampered by the required monetary and time commitment. Therefore, representative benchmarks are often chosen to estimate the performance of the actual application in advance (\eg in~\cite{Mao:2012gf,Schad:2010uq}).

Systematic cloud benchmarking is an elaborate task and demands for automation to efficiently conduct various benchmarks. Although representative benchmarks are typically much easier to deploy and execute on cloud services than actual applications, testing multiple providers with variable configurations results in a large parameter space to explore, making this kind of benchmarking still labor intensive. Moreover, in fast moving cloud environments, continuous reevaluation is inevitable, when providers change their supported instance types or upgrade their hardware. Therefore, several research projects \cite{Cunha:2013wu,Jayasinghe:2013fk,Silva:2013hb} aiming at extensible cloud benchmark automation were recently introduced. They all facilitate systematic cloud benchmarking, however, defining benchmarks is typically a tedious and error-prone activity. It often involves manually creating VM images for each benchmark, cloud provider, and region. This increases the time necessary to benchmark a given configuration, and reduces comparability and reproducability of results.  Time-consuming benchmark preparation was identified as recurring problem especially for application benchmarks in \cite{Cecchet:2011pd} and application deployment in general was mentioned as a key challenge for cloud computing in \cite{Moreno-Vozmediano:2013dq}. These challenges led us to the formulation of the following research questions, which guide this paper.

\begin{itemize}
  \item \textbf{RQ1:} How can existing IaaS cloud benchmarks be described in a  modular and portable manner?
  \item  \textbf{RQ2:} How can such benchmarks be periodically scheduled and executed in cloud environments in a fully reproducible way, and without manual interaction?
\end{itemize}

We argue that a suitable answer to these questions is to adopt the notion of Infrastructure-as-Code (IaC) for benchmarking, as introduced by the current DevOps trend~\cite{huettermann:12}. In IaC, the complete provisioning and configuration of various middleware components, most importantly IaaS VMs, operating systems, and standard software, is captured in provisioning code. Applying provisioning code reproducibly converges a system to a desired state, without the need for manual steps and irrespective of previous configurations of the same components. This concept is known as idempotence~\cite{hummer:13}.

We present the Cloud WorkBench (\cwb) framework, which is grounded on IaC to foster simple definition, execution, and repetition of benchmarks over a wide array of cloud providers and configurations. Further, we give an example of the capabilities of \cwb by using it to analyse the disk I/O speed of various standard VM instance types in Amazon EC2. Disk I/O speed is one of the most relevant indicators for estimating the performance of On-Line Transaction Processing (OLTP) applications in the cloud. 

The rest of this paper is structured as follows. We introduce \cwb in Section~\ref{sec:arch}. In Section~\ref{sec:case}, we illustrate the possibilities enabled by \cwb with a case study. We compare \cwb to the current state of research in Section~\ref{sec:rw}. Finally, in Section~\ref{sec:conc}, we conclude the paper with an outlook on future work.

\section{Cloud WorkBench Architecture}
\label{sec:arch}

This section introduces the \cwb framework for defining, scheduling, and executing benchmarks.

\subsection{System Overview}

\begin{figure}[h!]
	\centering
	\includegraphics[width=0.5\textwidth]{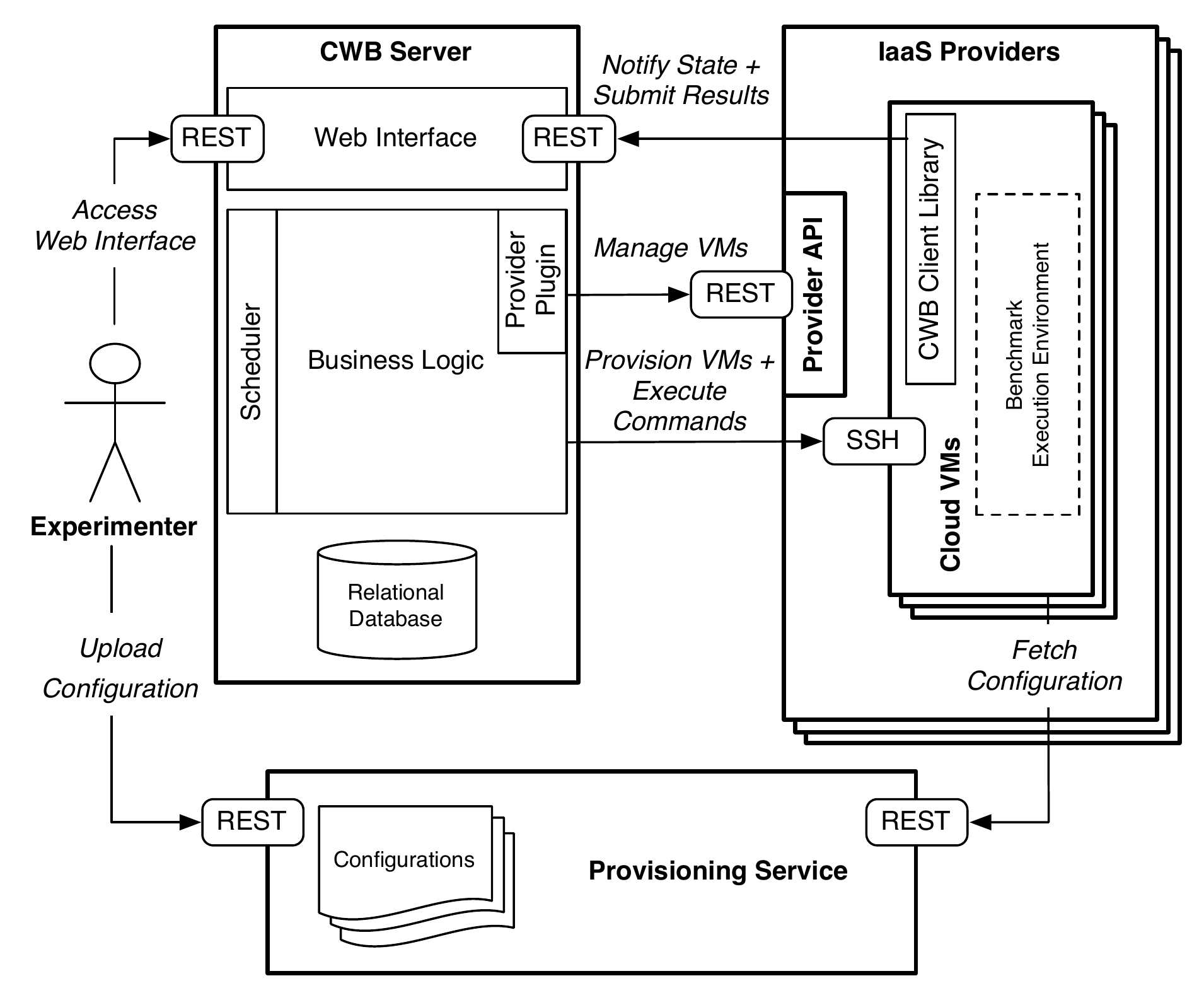}
	\caption{Architecture Overview}
	\label{fig:architecture-overview}
\end{figure}

Defining and executing a benchmark in \cwb involves interactions among five main components, as illustrated in Figure~\ref{fig:architecture-overview}.

The (human) \textit{experimenter} defines benchmarks via the provisioning service and the \cwb web interface, which subsequently allows one to schedule and manage executions of benchmarks.
The \textit{\cwb server} is the main component of the system, consisting of a standard three-tier web application. It provides the web interface, implements the business logic in collaboration with external dependencies, and stores its data (benchmark definitions and benchmark results) in a relational database. A component of the \textit{\cwb server} business logic is the scheduler, which periodically executes the defined benchmarks.

Benchmarks in \cwb are typically defined across a multitude of different IaaS providers, which the \textit{\cwb server} interacts with over a \textit{provider API}. Fundamentally, this API is mostly used to acquire and release cloud VMs of a given user-defined specification. These \textit{cloud VMs} are the System Under Test (SUT) and execute the actual benchmarking code. To ease the interaction between the \textit{cloud VMs} and the \textit{\cwb server}, a small \cwb client library is installed in each VM. This client library, along with all other required code (\eg Linux packages required by a benchmark, or the benchmark code itself), is provisioned in the \textit{cloud VMs} based on IaC configurations retrieved from a \textit{provisioning service}. The \textit{provisioning service} knows how to prepare a given bare VM to execute a given benchmark.

All interactions among these components happen typically over REST services to foster loose coupling and reusability, with the exception of the interaction between the \textit{\cwb server} and the \textit{cloud VMs}. These components communicate over the standard Linux utilities \texttt{rsync} and \texttt{ssh} for reasons of simplicity.

\subsection{Benchmark Definition}

\begin{figure}[h!]
	\centering
	\includegraphics[width=0.35\textwidth]{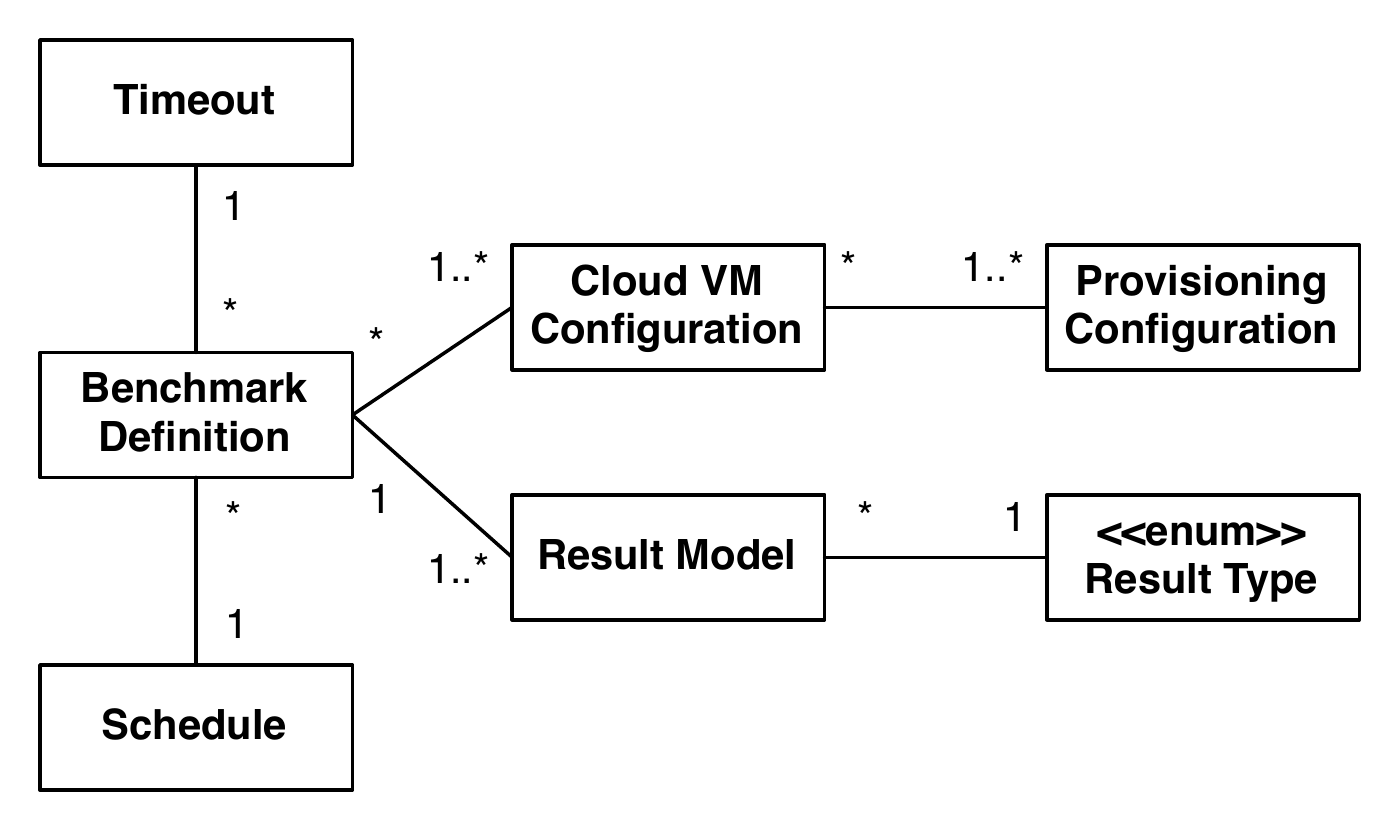}
	\caption{Structure of a Benchmark Definition}
	\label{fig:benchmark}
\end{figure}

One core feature of \cwb is that benchmarks, including the cloud configuration they are evaluating, can be defined entirely in code and by using the \cwb web interface, essentially following the ideas of DevOps and IaC. As argued in~\cite{huettermann:12}, this renders the process reproducible, modularizable, flexible, and testable using standard software engineering techniques. Common components among benchmarks can be easily shared and provisioning configurations from a large provisioning service community can be reused to efficiently describe the benchmark installation. 
Logically, a benchmark definition requires the information depicted in the simplified UML class diagram in Figure~\ref{fig:benchmark}.

Every benchmark definition requires one or more client VMs, which are brought into the expected configuration state via executing one or more provisioning configurations. Both, the definition of client VMs and provisioning configurations, follows the established notions of standard IaC tooling, \eg Vagrant\footnote{\url{http://www.vagrantup.com}} and Opscode Chef\footnote{\url{http://www.getchef.com}}. In addition, every benchmark definition requires one or more result models, which capture the type of outcome a benchmark will deliver. Finally, benchmarks optionally also contain a schedule (benchmarks without a schedule are only triggered manually by the experimenter) and a timeout, after which the execution of a benchmark is terminated no matter whether it is finished or not.


\cwb defines an interface to handle the interaction with user-defined benchmarks. Each benchmark must implement a callback (\ie a piece of code following a defined convention, which can be invoked by the \cwb server) to start executing. Further, benchmarks should use the provided \cwb client library to notify state updates (\eg when the benchmark run is completed or a failure has occurred, more details on the \cwb state model will follow in Section~\ref{sec:states}) and submit results back to the \cwb server. 
The client library can easily be installed via a pre-defined provisioning configuration. One additional advantage of this benchmark definition model is that experimenters can easily define variations of the same benchmark, \eg to execute identical benchmarks against a number of different cloud VM configurations. This is facilitated by \cwb via features for cloning and modifying existing benchmark definitions. As the provisioning code is logically separated from the definition of the cloud VMs, it is hence easy to set up a large array of benchmarks that evaluate different cloud configurations, and be confident that the code and setup of each benchmark is in fact identical except for the facets that the experimenter specifically wants to vary.


\subsection{Executing Benchmarks}

\begin{figure}
	\centering
	\includegraphics[width=0.5\textwidth]{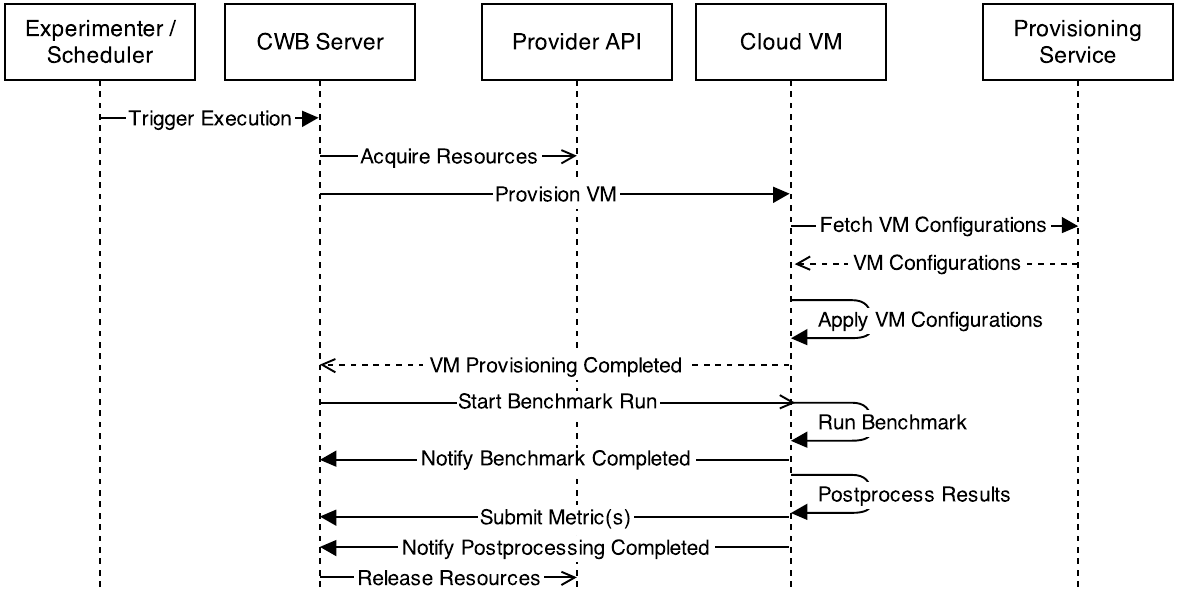}
	\caption{Executing a Benchmark}
	\label{fig:runtime-view}
\end{figure}

Figure~\ref{fig:runtime-view} illustrates the interactions when a new benchmark execution is triggered by the experimenter or the scheduler. For simplicity, we focus on a successful execution here (\ie neither provisioning nor benchmark execution  fails, and the benchmark finishes before the defined timeout is exceeded). Firstly, a provider plugin in the business logic asynchronously acquires resources (typically cloud VMs, but may also comprise cloud specific features, such as dedicated block storage or dynamically mapped IP addresses). As soon as the business logic successfully managed to establish a remote shell connection to the cloud VM, it starts orchestrating the VM provisioning via the remote shell connection. Thereby, each cloud VM fetches its role-dependent configurations from the provisioning service and applies them. At this point, the benchmark is entirely prepared for execution and asynchronously started via a remote shell command. This command invokes a defined callback on the VM that any benchmark has to implement. Once the actual benchmark workload is completed, the benchmark should notify this state update to the \cwb server via the client library. The benchmark results are then postprocessed, which typically involves textual result extraction, and submitted to the \cwb server as individual metrics or as a collection of metrics via a CSV file. After completed work, the cloud VM notifies the state update to the \cwb server to trigger all resources being released.

\subsection{Benchmark State Model}
\label{sec:states}



\begin{figure}
	\centering
	\includegraphics[width=0.38\textwidth]{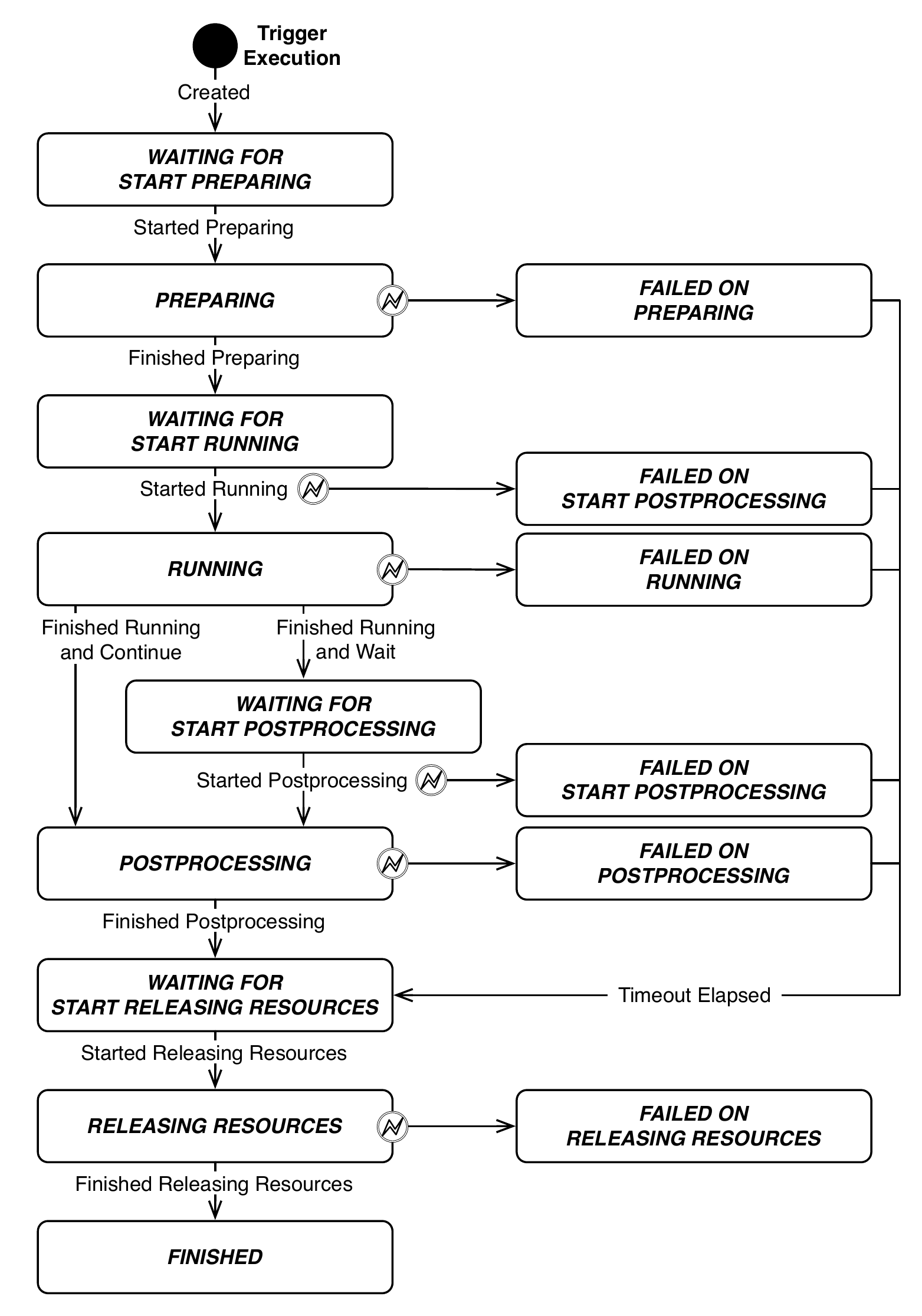}
	\caption{Benchmark Execution State Model}
	\label{fig:state-model}
\end{figure}

Every benchmark execution in \cwb runs through a non-trivial state model during their lifetime. Changes in the state model are triggered either by the \cwb server (for provisioning or resource cleanup related states) or by the benchmark itself via the client library (for execution-related states). Each state change is associated with a given event (\eg \textit{created}). Figure~\ref{fig:state-model} presents the benchmark execution state model.

An execution is \textit{created} either manually by the experimenter via the web interface, or automatically via a schedule. It is then \textit{WAITING FOR START PREPARING} until the \cwb server has processing capabilities available to start preparation (i.e., acquiring the VMs from the cloud, and applying provisioning code). Immediately before starting preparation, the event \textit{started preparing} is fired and subsequently, during preparation, the execution is in the \textit{PREPARING} state. Unhandled exceptions during preparation (\eg if the requested VM fails to launch) cause the execution to enter the \textit{FAILED ON PREPARING} state and release the acquired resources after a configurable timeout has elapsed. This timeout gives the experimenter the opportunity to activate the interactive development mode, fixing any provisioning errors, and reprovision the cloud VMs again. Interactive development mode introduces additional events and states that are not covered here as they are mostly relevant during development and testing of a benchmark. After an execution has \textit{finished preparing}, it is \textit{WAITING FOR START RUNNING} until the \cwb server has free processing capabilities available. The benchmark run is then started on the cloud VM via a remote shell command resulting in a \textit{started running} event in case of success and \textit{failed on running} event and state on failure. Failures on start running and failures described in the following are treated the same way as failures on preparing, that is the acquired resources are released after a timeout has been elapsed.

A successfully started benchmark is \textit{RUNNING} until a cloud VM notifies its completion, or the specified timeout from the benchmark definition has been elapsed. In the latter case, the execution is being treated as \textit{FAILED ON RUNNING}, as it failed to complete within the expected time duration. The \textit{FAILED ON RUNNING} state can also be reached if a cloud VM detects and notifies a failure by itself, for instance by catching a fatal error during benchmark execution. After \textit{finished running}, a cloud VM may either immediately continue with postprocessing, or enter the \textit{WAITING FOR START POSTPROCESSING} state until the \cwb server has processing capabilities available to trigger postprocessing. This indirection is aimed to support multi-VM benchmarks, where the responsibilities for recognizing benchmark completion and postprocessing are taken by distinctive cloud VMs. Otherwise, postprocessing follows the pattern for asynchronously executed remote commands (\eg running the benchmark) and releasing resources follows the pattern for locally executed commands (\eg preparing the benchmark). Errors while releasing resources are particularly problematic, as those may leave back costly residual cloud VMs, which need to be destroyed manually by the experimenter. All other error cases do not have this problem, as \cwb will always make sure to reach a clean state (\ie all resources are released) after the execution timeout of a benchmark is elapsed.
Executions without any failures remain in the \textit{FINISHED} state after having \textit{finished releasing resources}. Executions that exhibit at least one failure show their first failure state so that the experimenter can easily recognize at what step an execution has failed.

\subsection{Benchmarking Results}
\label{sec:results}

The observed results of a benchmark execution are represented differently based on the type of their definition. Currently, \cwb supports four common types of results (nominal, ordinal, interval, and ratio scale, following \cite{Stevens:1946kx}). Nominal scale results are stored as String data types whereas other scale types are represented as floating point data types. This distinction enables efficient sorting at database level whereas presenting a uniform interface to the rest of the application by abstracting implementation details. The scale type of results also has implications for any data analysis based on the collected benchmark results.

\subsection{System Implementation}
\label{sec:impl}

The \cwb web application is implemented using the Ruby on Rails\footnote{\url{http://rubyonrails.org/}} framework. One of the fundamental distinctions between \cwb and related work is that we strived to reuse as much existing DevOps tooling as possible, so that experimenters can build upon existing community artifacts (e.g., for provisioning configurations) and knowledge.
Hence, we use Cron as scheduler, Vagrant as VM environment management tool, and Opscode Chef as a provisioning tool. 

Vagrant was chosen to represent cloud VM configurations using an established Ruby-based DSL. It abstracts cloud provider APIs, provisioning orchestration, and the execution of remote shell commands. The DSL exposes all relevant configuration options in a declarative and easy-to-understand manner. Vagrant provides open source plugins for all relevant IaaS providers. The \cwb web interface integrates a minimal web IDE with syntax highlighting for the Vagrant DSL.

Choosing Opscode Chef with a dedicated Chef server as provisioning service provides us a flexible way to install and configure benchmark components in a reusable manner by exploiting Chef attributes. Experimenters can reuse software components (\eg database installation and setup code) in terms of cookbooks from the worldwide Chef community, and easily share benchmark infrastructure code with others. Furthermore, Chef integrates particularly well with Vagrant. The attribute passing mechanism from Vagrant to Chef allows to build configurable and thus reusable benchmark cookbooks. Since both, Chef and Vagrant, use an internal Ruby DSL, they not only ensure language consistency across the project but also offer the capabilities of a fully featured programming language that is exploited with the use of variables and utility functions.

\begin{figure}[h]
	\centering
	\includegraphics[width=0.5\textwidth]{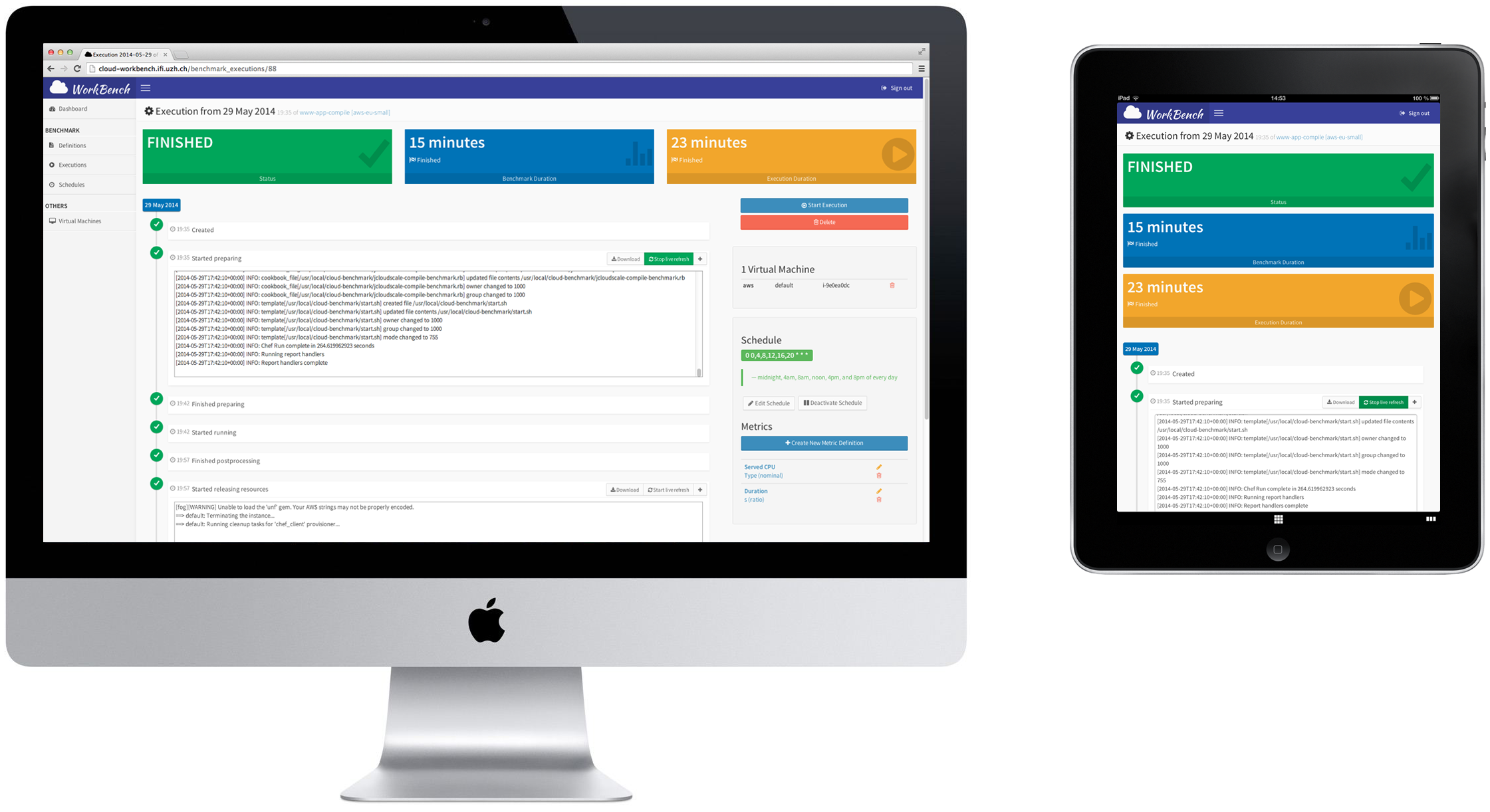}
	\caption{Responsive Web Client}
	\label{fig:desktop-and-tablet}
\end{figure}

The web interface takes advantage of the popular Bootstrap\footnote{\url{http://getbootstrap.com/}} front-end framework and is visually enhanced using an open source template and custom styling. It provides basic CRUD operations for the application entities where meaningful, context dependent tabular listing of entities, some basic search and filter operations and live log refresh via Ajax. Figure~\ref{fig:desktop-and-tablet} exemplifies its ability to adapt to different types of devices by dynamically rearranging the user interface elements appropriately.

The current version of \cwb is available as an open source project from Github\footnote{\url{https://github.com/sealuzh/cloud-workbench}}, including samples and installation instructions. The \cwb web interface itself can easily be set up in an IaaS cloud using Vagrant.

\section{Case Study}
\label{sec:case}

To illustrate the capabilities of \cwb, we now present a small-scale experiment regarding \textit{sequential disk write speed} in Amazon EC2. We aimed to answer the following concrete questions, which could in a similar way also be asked by a practitioner aiming to deploy an IO-intensive application, e.g., a database system, on EC2: \emph{(1) When is the sequential disk write speed of larger instance types better than of smaller instance types? (2) When should larger instance types be preferred over the better block storage type? (3) How do instance types and block storage types influence the variability of the sequential disk write speed?}

\subsection{Study Setup}

The data for this study was collected between June 20th and 23th, 2014 distributed over the day. Experiments were repeated for each setting 8 to 12 times depending on the observed variability. All experiments were conducted in the EC2 region Ireland (\texttt{eu-west-1}) using Ubuntu 14.04 images. We conducted our experiments on three different instance types (\texttt{t1.micro}, \texttt{m1.small}, and \texttt{m3.medium}). Additionally, 20 GB of Amazon EBS was provisioned for each instance. For this additional storage, we chose either the magnetic volumes or the newer (and, at the time of this writing, twice as expensive) general-purpose SSD EBS. This information was captured in Vagrantfiles via the \cwb web interface. We used the FIO\footnote{\url{http://git.kernel.org/cgit/linux/kernel/git/axboe/fio.git}} 2.1.10 benchmark. The sequential write is performed with workloads of 1 GiB ($\sim$1074 MB) and 4 GiB ($\sim$4295 MB) using the default block size of 4 KiB (4096 bytes). Direct I/O mode is used to assess the raw write performance ignoring caches. Additionally, the refill buffers mode is enabled to prevent SSD compression effects.

\begin{verbatim}
# Update package index
include_recipe "apt"

# Install the FIO benchmark
# via package manager
package "fio"
\end{verbatim}

A Chef cookbook was created that describes the FIO benchmark in a configurable manner. The listing above shows the part of the Chef cookbook being responsible for benchmark installation. Evidently, the fact that \cwb builds on top of Chef makes this step trivial for the benchmark developer. This cookbook also generates Ruby code to start the execution, postprocess the results, submit the observed metrics, and notify state updates to the \cwb server. Thereby, two metrics are defined and reported. Firstly, the CPU model name and secondly, a log of the bandwidth with the resolution of 500 milliseconds.

\begin{verbatim}
Vagrant.configure("2") do |config|
  config.vm.provider :aws do
    |aws, override|
    aws.region = "eu-west-1"
    aws.ami = "ami-896c96fe"
    override.ssh.username = "ubuntu"
    aws.instance_type = "m1.small"
  end

  config.vm.provision "chef_client" do
  |chef|
    chef.add_recipe
      "recipe[fio-benchmark@0.3.0]"
    chef.json =
    {
      "fio" => {
        "metric_definition_id"
          => "seq. write",
        "config" => {
          "size" => "1g",
          "refill_buffers" => "1"
        }
      }
    }
  end
end
\end{verbatim}

The Vagrantfile above then specifies the desired cloud resources, references the FIO benchmark Chef cookbook, and passes optional configuration.


\subsection{Results and Discussion}

Next, we briefly discuss the answers we obtained from the questions raised at the beginning of this section.

\subsubsection{Difference in Write Performance of Different Instance Sizes}

\begin{figure}[h]
	\centering
	\includegraphics[width=0.4\textwidth]{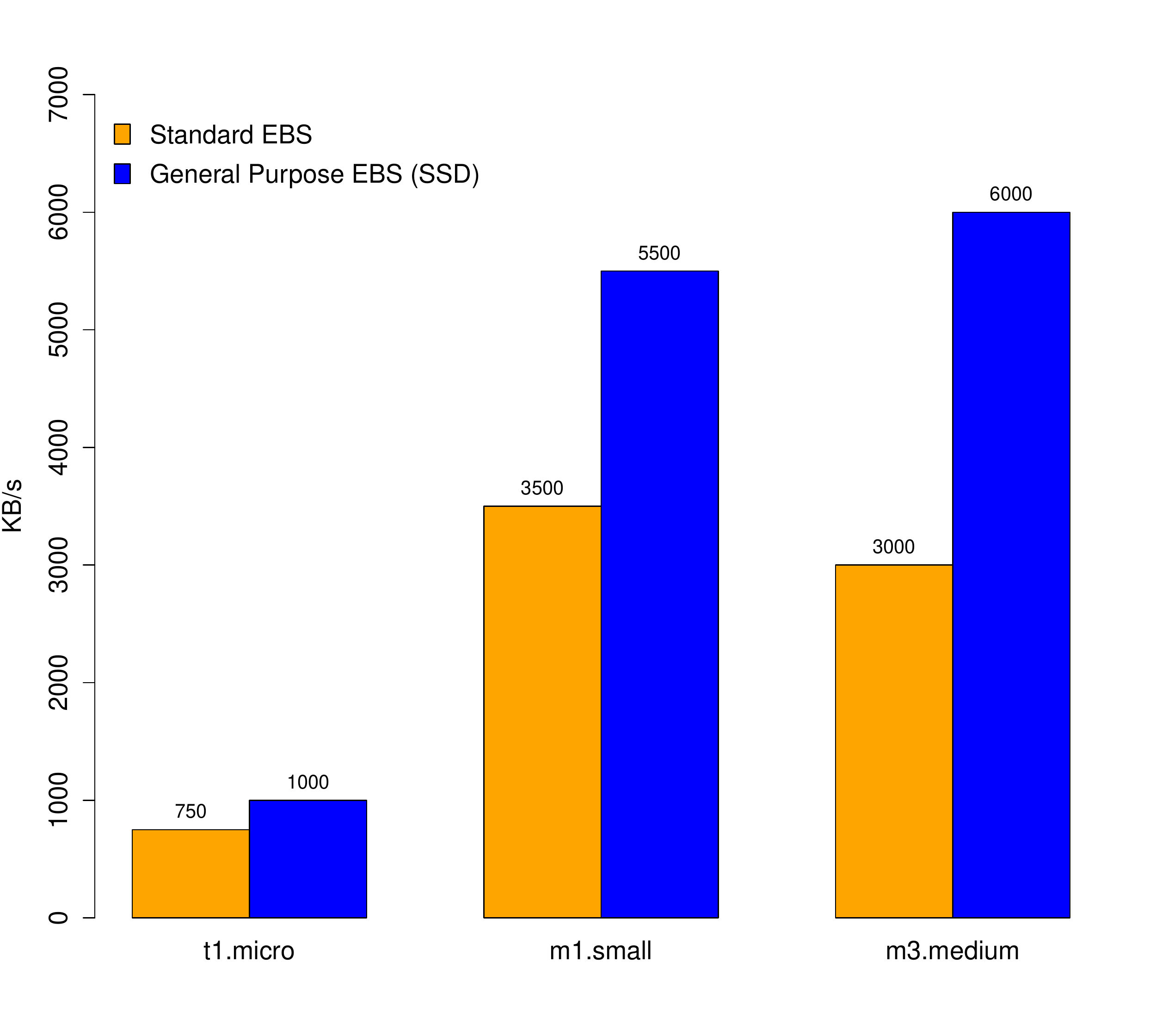}
	\caption{Sequential Write Bandwidth by Instance and Storage Type}
	\label{fig:instance-types}
\end{figure}

Raw sequential write performance increases by about factor 4 for both EBS types when upgrading from the smallest instance type in the study (\texttt{t1.micro}) to the next larger instance type \texttt{m1.small} or the even larger instance type \texttt{m3.medium}. Figure~\ref{fig:instance-types} illustrates this performance increase, but also reveals that larger instance types do not necessarily perform better than smaller instance types in all cases.

The large difference between \texttt{t1.micro} and the larger instance types may be explained with resource sharing and limited networking capabilities of the \texttt{t1.micro} instance type. Similar to other newer bursting instance types, \texttt{t1.micro} shares its single CPU with another tenant. Therefore, it only gets half of the CPU cycles at maximum \cite{Wang:2010ix}, which may affect the disk I/O performance \cite{Gregg:2013ys}. Networking performance influences the disk I/O performance, as block storage is connected to the VM instance as network storage. The networking performance specification provided by Amazon is very vague. Thus, a potentially significantly slower networking performance of \texttt{t1.micro} may degrade its disk I/O performance. In order to assess these assumptions, further studies may correlate disk I/O performance with CPU and networking performance.


\subsubsection{Contrasting Different Instance Types and Different Storage Options}


Larger instance types should be preferred over the better block storage type when using a \texttt{t1.micro} instance type. This is shown by Figure~\ref{fig:instance-types}, as the absolute performance gain is much higher when upgrading from \texttt{t1.micro} to \texttt{m1.small} (+2750 KB/s) than when upgrading from the standard to the general purpose storage type (+250 KB/s). The combination of \texttt{m1.small} with standard EBS will be more expensive than \texttt{t1.micro} with general purpose EBS for block storage sizes below 350 GBs. However, without considering I/O operation expenses, the cost/performance ratio is always better for the \texttt{m1.small} and standard EBS combination. Disk I/O intensive applications with more than one million I/O requests per hour can shift this ratio in favor of the other combination (\texttt{t1.micro} with general purpose EBS).

On the contrary, the better block storage type should be preferred over larger instance types when using a \texttt{m1.small} instance type. General purpose EBS can improve the performance of a \texttt{m1.small} instance type while the performance remains approximately the same when upgrading to the larger \texttt{m3.medium} instance type. Additionally, the cost/performance ratio is always better with the \texttt{m1.small} and general purpose EBS combination. Considering the expenses for I/O requests will even increase this advantage.

These results indicate that standard EBS is limited to approximately 3500 KB/s whereas the performance of the \texttt{t1.micro} instance type is restricted for other reasons. Similarly, general purpose EBS reaches approximately 6000 KB/s with the two larger instance types whereas its performance is restricted to about 1000 KB/s by the \texttt{t1.micro} instance type.

\subsubsection{Performance Variability of Instance and Block Storage Types}


\begin{table*}[t!]
\centering
\scriptsize
\begin{threeparttable}
  \caption{Sequential Write Bandwidth Variability (1 GiB)}
	\label{tab:variability}
	\centering
  \begin{tabular}{| p{4.5cm} | p{2.3cm} | p{2.3cm} | p{2.3cm} |}
  \hline
   & \texttt{t1.micro} & \texttt{m1.small} & \texttt{m3.medium} \\ \hline
  Standard EBS & 20\% (20-50\%) & 20\% (10-20\%) & 30\% (15-60\%) \\ \hline
  General Purpose SSD EBS & 10\% (20-40\%) & 10\% (5-15\%) & 10\% (5-10\%) \\ \hline
  \end{tabular}
     \begin{tablenotes}
      \small
      \item The variability is given as standard deviation in percentage of the mean across and within (in brackets) distinct benchmark executions.
    \end{tablenotes}
    \end{threeparttable}
\end{table*}

In general, the observed disk I/O performance varies remarkably. However, two patterns were recognized when comparing the variability across and within distinct benchmark executions for different types of VM instances and EBS. Firstly, standard EBS exhibits larger variability than general purpose SSD EBS for all instance types across and within distinct benchmark executions, as shown by Table~\ref{tab:variability}. Secondly, the \texttt{t1.micro} instance type exhibits much larger variability within, but not across, distinct benchmark executions compared to the larger instance types. Table~\ref{tab:variability} shows this unusually high performance variability of 20 to 50 \% for both EBS types within single benchmark executions for the \texttt{t1.micro} instance type.

The fact that this extraordinary high variability is equalized across distinct benchmark executions supports the assumption that CPU scheduling negatively influences the performance, as the CPU scheduling effect is only recognizable in the high resolution performance analysis conducted within single benchmark executions.

\begin{figure}
	\centering
	\includegraphics[width=0.45\textwidth]{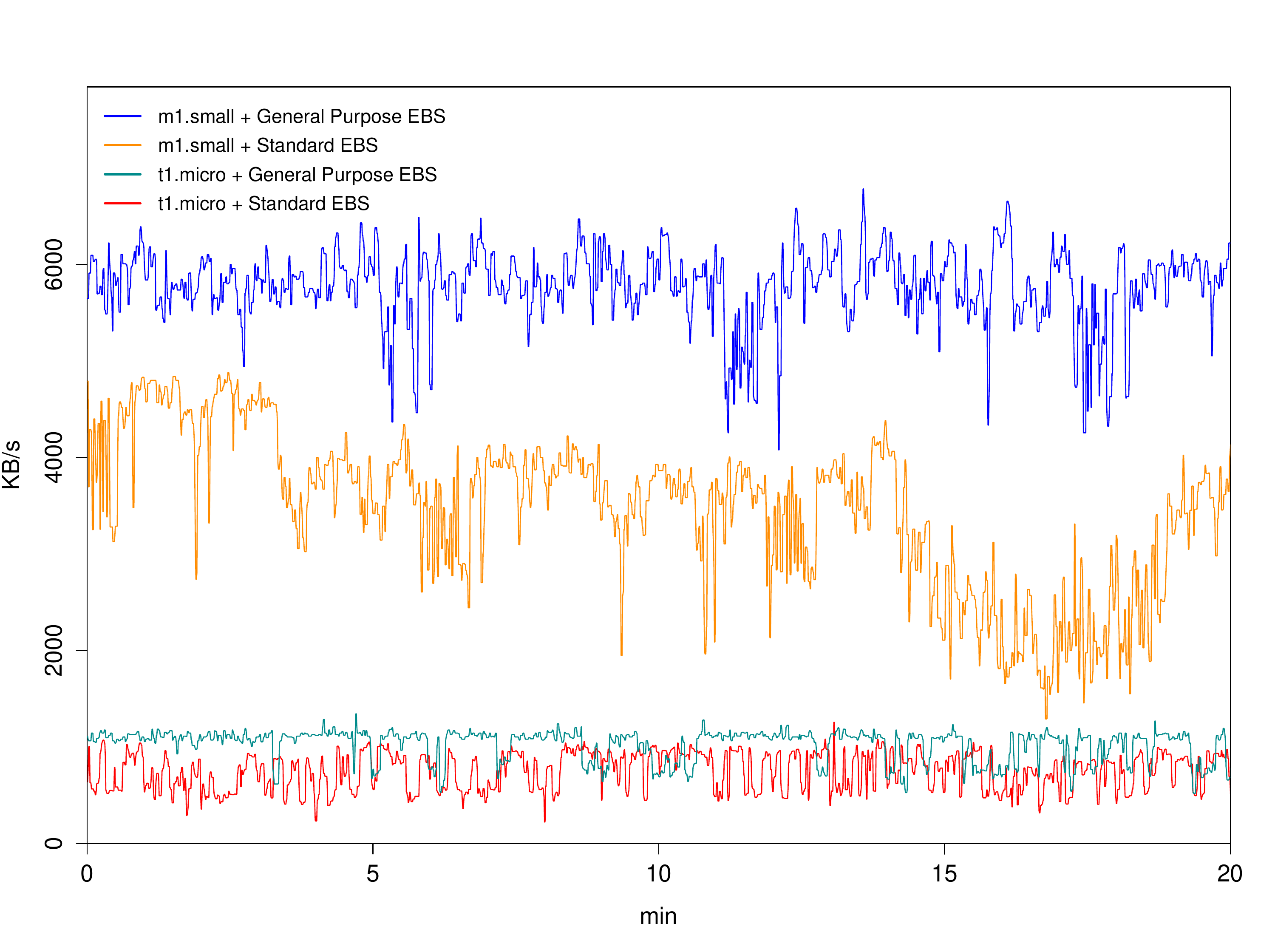}
	\caption{Bandwidth of Single Executions over Time}
	\label{fig:variability}
\end{figure}

Variability within a single execution is mostly ignored in literature, and only the average value of single executions are collected. Although this makes sense in general, analyzing single executions in detail can help to better understand the nature of disk I/O performance. Figure~\ref{fig:variability} compares the sequential write performance over time for single executions of the instances types \texttt{t1.micro} and \texttt{m1.small} in combination with standard and general purpose EBS. It illustrates strong oscillation for both instance and storage types. This common behavior appears in combination with sudden performance drops that may turn out even stronger and endure even longer than illustrated by the curve \texttt{m1.small} with standard EBS especially around minute 15. In addition, this curve exemplifies the unpredictable performance behavior of standard EBS exhibiting arbitrary ups and downs. On the contrary, the bandwidth for general purpose SSD EBS typically oscillates around the mean but still periodically drops in performance.

\section{Related Work}
\label{sec:rw}

The need for supporting experimenters in conducting benchmarks in cloud environments has already been recognized in existing literature. Studies whose goals most closely match those of \cwb include CloudBench \cite{Silva:2013hb}, Expertus \cite{Jayasinghe:2013fk,Jayasinghe:2012fk}, and Cloud Crawler \cite{Cunha:2013wu}. 

CloudBench \cite{Silva:2013hb} proposes a mainly imperative approach \cite{Cunha:2013wu} for defining benchmarks at different levels of abstraction. New benchmarks are defined via a high-level experiment plan, one or multiple medium-level application or workload templates, and multiple low-level hook shell scripts. The authors claim that CloudBench can \emph{"represent and benchmark almost every observable interaction of a cloud"} \cite{Silva:2013hb}. Comprehensive literature review has shown that CloudBench is currently the most sophisticated and extensive approach. Especially its capabilities to execute complex and dynamic scale-out benchmarks as demanded in \cite{Binnig:2009fk,Cooper:2010vn,Ferdman:2012fk} makes CloudBench unique.
Expertus, introduced in \cite{Jayasinghe:2012fk} and extended in \cite{Jayasinghe:2013fk}, proposes a code generation based approach for defining benchmarks. New benchmarks are defined via XSLT templates that generate shell scripts for the individual cloud VMs. Expertus further provides benchmark and workload configurability via XML, large-scale metric collection and a web interface with interactive visualization and statistical analysis capabilities.
Cloud Crawler \cite{Cunha:2013wu} proposes a declarative approach for defining benchmarks. New benchmarks are defined via its own external domain-specific language using a YAML-based syntax. Additionally, each benchmark must extend the Crawler execution engine by implementing a Java interface.

\cwb differentiates from these previous approaches via its strong IaC core, which makes it easy to define benchmarks based on standard tooling and concepts, as well as share benchmark definitions. Further, none of the previously described approaches is known to offer provisioning capabilities to the same extent and with the same modularity as \cwb does. There is also no solution known that integrates periodic scheduling functionality into a web-based framework. Furthermore, \cwb together with CloudBench are the only frameworks designed for benchmark extensibility at runtime.

\section{Conclusions}
\label{sec:conc}

This paper presented a web-based framework called \cwb, which supports experimenters in conducting IaaS cloud benchmarks. \cwb was designed and implemented to leverage the notion of IaC for cloud benchmarking, and is used to automate the benchmarking lifecycle from the definition to the execution of benchmarks.
Currently, we are using \cwb to execute extensive benchmarks over different cloud providers. At the time of this writing, we have already collected data for close to 20000 benchmark executions using our \cwb tooling, illustrating the system's suitability of real-life use.

As part of our future work, we plan to add support for additional cloud providers, automate the collection of common metrics, integrate statistical analysis and visualization capabilities, and facilitate benchmark definition even more. The ultimate goal of \cwb is to support the entire benchmarking lifecycle, from benchmark definition to the statistical analysis and visualization of the observed metrics, via a single web-based toolkit.

\section*{Acknowledgements}
The research leading to these results has received funding from the
European Community's Seventh Framework Programme (FP7/2007-2013) under grant
agreement no. 610802 (CloudWave).

\bibliographystyle{IEEEtranS}
\bibliography{bibtex}

\end{document}